\documentclass[aps,pra,onecolumn,nofootinbib,superscriptaddress]{revtex4-2} 
\usepackage[normalem]{ulem}

\begin{document}

\title{Integrating Across Application, Model, Algorithm, Compilation, and Error Correction Chasms With \sout{QASM} Quantum Type Theory}
\author{Eugene Dumitrescu}
\email{edq@ornl.gov}
\affiliation{Quantum Information Science Section, Oak Ridge National Laboratory}

\maketitle

\textit{Challenge---} 
\textit{Efficiently} and \textit{correctly} programming a quantum computer is difficult in part because, by connecting high-level algorithmic constructs with a physical-level hardware realization, a complete quantum computation traverses wide chasms between vastly different hierarchical computational scales. As the technical machinery to express even a simple quantum computation across \textit{all} scales is lacking, we propose type theory as a central ingredient in efficient and correct quantum programs. 
Different relevant (as well as marginal and irrelevant) types, associated with instructions operating upon them, should come into focus as one re-scales along the physical- to algorithmic-levels in such a consistent hierarchical type theory. 
Likewise, an understanding all hierarchical levels and their properties is required for a true end-to-end compiler.  

For any type to be useful a working standard operating library is also required. Standard libraries defining arithmetic operations acting on quantum types (qint, qfloat, and qetc defined later), as well as time-evolution protocols, and standard oracular gadgetry are required to construct high-level programs. Some type properties are transferable across levels, while others are not. To unify this, a core representation must exist which expresses salient attributes at each hierarchical level. While some software has been dedicated to investigating this interplay of types and data-structures \cite{McClean2020, Mintz2020}, more are needed to deliver complete type theory which is fully operational and correct (whether by proof or by unit-test). While library constructions will require great efforts, standardizing types will dramatically reduce the effort to maintain codes as well to develop a myriad of new languages, compilers, and programs. 

\textit{Opportunity---}  
To highlight and organize type theory's implications across the spectrum we add a new attribute to quantum types. That is, define a \textit{color} quantum attribute number $\mu \in \{r,g,b\}$ denoting the quantum computing types' and instructions' levels within the physical-to-algorithmic computational hierarchy. The largest-scale, so called infrared-red (IR), scale logical operational modalities are red (r), error correction coarse-grains through the intermediary green (g) spectrum, and the microscopic physical-length-scale noisy ultra-violet (UV) types and instructions are designated blue (b). Analog (sometimes called noisy intermediate scale quantum) experiments reside in a limited near-term computational modality where this color hierarchy collapses ($\mu = r = g = b$) which indicates that the types and instruction sets used across the hardware and the logical levels are identical. As error correction first emerges there will be, for the first time, a distinction between the physical and logical types. As outlined below, we argue that realizing universal quantum computations will be composed of types and compilers throughout the spectrum. 

\textit{Assessment---} At the different layers of a quantum computation:

\textit{(r) Science Oriented Languages---} From a \textit{domain scientist's} perspective high-level quantum type theory is the crucial tool to formulate, compile, and compute specific problems of interest. The working domain scientist already implicitly has a good grasp of quantum types since they previously utilized classical instantiating (and approximations) of fundamental quantum types in modeling physical phenomena. Here one utilized the familiar and expressive classical types (bits, bytes, ints, floats, chars, classes, etc) defined within the context of a given programming language to models properties of, for example, quantum fields. 

Already at this level, a need for fundamental quantum types for expressive and operational purposes emerges. Fundamental quantum computing types should include qubit, fermion, qumode, ancilla, interferometer, (weak) measurement, and classical groups (and eventually beyond). Such quantum types are needed to both express generic as well as design optimal quantum algorithms. In a Feynman-like approach, simulating fermions with a fermionic quantum computer \cite{Bravyi2002} (equipped with type $f_r$) is the most straightforward path for the high-level chemist. However due to the difficulties of simulating fermions, as well as the needs for other types, one often begins with other types such as $q_r$, $b_r$, $a_r$ for qubit (represented by Pauli-algebra), boson (canonical commutation relations), and anyon (defined by the parent quantum field theory's topological datum \cite{Nayak2008,Johansen2020}) respectively. 

\textit{(g) Error Correction---}
Along with lowering the noise floor, error correction is the substrate mapping, distributing, and enabling a manifestly digitized quantum communication and computation. The latter component is of special interest because it maps types and instructions between the bare and encoded layers. For example, superconducting and atomic qubits have both been used to mimic small scale quantum error correcting codes. Hence we have a first example of two experiments which are fundamentally distinct at the $\mu = b$ physical-level but are logically equivalent in operation (up to error rates) using the encoded $g$ logical types. Note that within this framework, many logical concatenations could abstractly be encapsulated at the emergent $\mu = g$ level. For example, one may envision first using a $[[n,k,d]]$ encoding, mapping the $n$ physical to the $k$ logical qubits $q_b \rightarrow q_{g_1}$. We could further concatenate codes to re-encode qubits, with e.g., a $[[n'=k,k',d']]$ encoding: $q_{g_1} \rightarrow q_{g_2}$. By defining a code with manifestly fermionic characteristics (other characteristics would accompany other types of emergent fields) we alternatively have an encoding map: $q_{g_1}\rightarrow f_{g_2}$ \cite{Landahl2021}. One can continue in this fashion further encoding the fermions into a higher type $f_{g_2}\rightarrow q_{g_3}$, or alternatively as might be the case for chemistry applications, just accept this type as sufficiently encoded to be regarded, and compiled to, as a high-level logical type: $f_{g_2} \equiv f_r$.  Just like at the logical level, standard libraries are required for error correction. 

\textit{(b) Microscopic Device Description---}
At the UV, physical length scale, a device implements ``digital''---i.e. pre-specified and hopefully well characterized analog instructions---dynamics to realize (quasi-)isolated Schrodinger-equation dynamics on a qubit's physical Hilbert space. These physical operations form the basis of either error correcting instructions or analog operations. Perhaps due to the emergence of superconducting\cite{Cross2017, Smith2016}, atomic\cite{Ebadi2021}, and photonic qubits \cite{Bartolucci2021}  (defining discrete \textit{qudit} or continuous variable \textit{qumode} types) a large amount of recent attention has been dedicated to formalizing \textit{qubit} types which are various instances of $q_b$. Similar to how a qubit type is implicitly and then later explicitly defined by their $SU(2)$ operations and generating local Pauli algebra, a qumode type with certain instruction sets (linear optics and two-mode squeezers) could also concisely and formally be defined with data structure being the group $SU(1, 1)$ \cite{Shaterzadeh-Yazdi2008}. Other types can and should also be defined by a succinct mathematical representation. For example, qubits are defined by space and time coordinates as well as algebraic computational operations. Here, at the level of $SU(2)$, the Euler, ZYZ Cartan, and $U_3$ decompositions are all synonymous (although they may be generalized in different ways). In addition, there is a further need to generically use the fundamental types as units building \textit{compound} quantum types such as a qreg (a qubit array).

\textit{Timeliness and maturity---} 
Using the color type attribute we can now organize and connect programs along a common (color) axis. For example, the popular pythonic framework Qiskit compiles a user's programs, with types at this bottom $b$-level, into a json dictionary containing a valid OpenQASM2.0 \cite{Cross2017} compiled program. OpenQASM3.0 \cite{Cross2022} tentatively makes a step towards our definition by separating along coherent (purely quantum types) and longer, incoherent, timescales (with low-level classical types). Such considerations are required to understand the $b$ level and its interactions with $g$ and $r$. The quantum intermediate representation (QIR), and prior works, have introduced initial LLVM compatible quantum compiler tool chains \cite{McCaskey2018, Nguyen2022}. However the current QIR lives close to the physical layer and does not couple to the $(r,g)$ degrees of freedom.  As with all three levels, the compilation framework should abide by principles of, or systematically reflect, linear logic and relevant properties of quantum information (no cloning, reversibility for pure systems, or contractiveness of dissipative maps). In this way, proper type theory aids facilitates the extermination quantum bugs from our programs \cite{Luo2023}. Of course, this is in addition to type theory defining, informing, and creating the structure to operationally translate quantum computations across scales. 

\bibliographystyle{unsrt}
\bibliography{main}
\end{document}